# NON-EQUILIBRIUM MOTT TRANSITION IN A LATTICE OF BOSE-EINSTEIN CONDENSATES

J. Dziarmaga [1,2], A. Smerzi [1], W.H. Zurek [1], A.R. Bishop [1]

[1] *Los Alamos National Laboratory, Theory Division, Los Alamos, NM 87545, USA*

[2] *Instytut Fizyki Uniwersytetu Jagiellońskiego, Reymonta 4, 30-059 Kraków, Poland*

**Abstract**
We study the non-equilibrium dynamics of a zero temperature Mott insulator-superfluid quantum phase transition in a lattice of weakly coupled Bose-Einstein condensates. We show that crossing the critical point from the insulating to the superfluid phase at a finite rate results in a finite dispersion of current. Viceversa, crossing the critical point from the superfluid phase to the insulating one, results in a finite dispersion in the number of atoms per lattice site. We develop a microscopic dynamical model of the transition, and make quantitative predictions for realistic experiments.

## 1. Introduction

The relaxation time $\tau$ and the correlation length $\xi$ of thermal fluctuations diverge when the control parameter of a continuous phase transition $\epsilon = \frac{T - T_c}{T_c}$, which measures the distance from the critical point, tends to zero: $\xi \sim |\epsilon|^{-\nu}$ and $\tau \sim |\epsilon|^{-z\nu}$, with $z$ and $\nu$ critical exponents.

In an adiabatic, infinitesimally slow transition trough the critical point, the symmetry broken phase is described by a fully ordered complex order parameter. However, when a quench takes place at a finite rate, the system goes out of equilibrium some time before the transition because of the critical slowing down of the relaxation time of the system ($\tau \to \infty$ when $\epsilon \to 0$). As a result, the complex order parameter which emerges after the transition will assume different random phases in different domains of space. This leads to formation of topological defects - e.g. vortices - wherever the circulation of the phase around a closed loop happens to be nonzero.

This process is known as Kibble-Zurek mechanism (KZM) [1–3]. It is rather simple to estimate the size of the domains [2]. We consider a linear time dependence of the control parameter $\epsilon(t) \approx t/\tau_Q$. The transition rate is $r(t) = \dot{\epsilon}/\epsilon = 1/|t|$, and the system goes out of equilibrium at the time $-\hat{t}$ before the transition when $r(-\hat{t}) = \tau^{-1}(-\hat{t})$. After that time the state of the system essentially does not change until $+\hat{t}$, when the rate $r(t)$ becomes again equal to the relaxation rate, $r(+\hat{t}) = \tau^{-1}(+\hat{t})$. At $\hat{t} = \tau_Q^{z\nu/(1+z\nu)}$ fluctuations of the order parameter with wavelengths longer than $\hat{\xi} = \xi(\hat{t}) = \tau_Q^{\nu/(1+z\nu)}$ begin to grow exponentially while short wavelength fluctuations remain unchanged. The size of the correlated domains is given by $\hat{\xi}$ and the density of vortices by $1/\hat{f}\hat{\xi}^2$ where $f \sim 10$ in numerical experiments [7] as well as in analytic treatements of exactly soluble models [8].

Theoretical and experimental studies of KZM have concentrated so far on thermal continuous phase transitions with the dynamics of the order parameter governed by a phenomenological, irreversible time-dependent Ginzurg-Landau theory. Most attention has been devoted to the normal-superfluid transition in $^4He$ [5], $^3He$ [6], superconductors [9, 10, 4], and, more recently, dilute Bose-Einstein condensates [11]. To date, truly microscopic approaches have been too complicated to extract useful predictions.

A quench-induced *quantum* phase transition (QPT) at temperature $T = 0$ must be treated in a microscopic way. It is a common wisdom that some properties of a quantum transition can be obtained by an exact map from a thermodynamic transition [12]: the correlation length in the ground state of the quantum system scales like $\xi \sim |\epsilon|^{-\nu}$ and the gap between the ground state and the first excited state like $dE \sim |\epsilon|^{-z\nu}$. To study the dynamics of a quantum transition, which drives the system out of its ground state, we also need information about correlations in excited states, which is not provided by this map. As we will see below, the essence of the KZM, which is the competition between the transition rate and the timescale on which the system can react, is applicable to quantum transitions. However, the quantum scenario and, in particular, the interpretation of its results differ from the thermal case. The main reason is the reversibility of the quantum dynamics, as opposed to the dynamical irreversibility of the thermal critical dynamics.

In this paper, we study the appearance of a nonzero current while the system is undergoing a (zero-temperature) quantum phase transition from the insulating to the superfluid phase, providing details of a previously published study [13]. We also study the dynamics of the transition crossing the critical point in the opposite direction, namely from the superfluid to the insulator phase. When the transition is diabatic, this results in a finite dispersion in the number of atoms per site. We develop a microscopic, dynamical theory and suggest experiments to test our predictions.

## 2. Josephson junction arrays

We consider a dilute gas of $N$ atoms, trapped in a deep optical lattice. The lattice can be generated by counterpropagating lasers, which create a regular (possibly multidimensional) array of wells separated by potential barriers. The atoms are trapped on the bottom of each well, creating a lattice of weakly coupled Bose-Einstein condensates, the weak coupling being created by the tunneling of the atoms through the interwell barriers. When the height of the barrier is much larger than the "local" chemical potential of each condensate, the system can be described (in the "tight binding approximation") by the Boson Hubbard Model (BHM) Hamiltonian:

$$H = -\frac{\gamma}{2} \sum_{\langle k,l \rangle} a_k^\dagger a_l + \frac{g\beta}{2} \sum_k (a_k^\dagger a_k)^2 \,, \tag{1}$$

where $g = 4\pi a \hbar^2/m$ is interaction strength, $a$ is the interatomic s-wave scattering length and $m$ the mass of the atoms. The term proportional to $\gamma$ in the BHM describes the atomic tunneling between the wells, while the term proportional to $g\beta$ takes in account the interaction energy of the condensate atoms in each well. The coefficients $\beta$ and $\gamma$ are integrals of the single particle bound state wave functions localized in each well [14].

We first consider the case when the total number of atoms $N$ is commensurate with the number of lattice sites, so that the average number of atoms per site $n$ is an integer. The implications of a non-commensurate $N$ will be discussed in Section 1.5.2.

In the commensurate case the properties of the ground state of the BHM are governed by a dimensionless parameter

$$G \equiv \frac{n\gamma}{g\beta} \tag{2}$$

It is possible to recognize four different regimes:

- $n^2 \ll G$ — the average interaction energy $ng\beta$ per atom can be neglected as compared to the hopping rate. The interaction term in the BHM (1) can be ignored and the ground state can be approximated by a quasi-coherent state

$$|GS\rangle \sim \left( \sum_k a_k^\dagger \right)^N |0\rangle \tag{3}$$

with the dispersion of the number of atoms in each site $\Delta n \sim n^{1/2}$, and almost vanishing dispersion of the phase difference between nearest neighbor sites,

$$\Delta\phi \sim \frac{1}{n^{1/2}} \ . \qquad (4)$$

- $1 \ll G \ll n^2$ —- The on-site energy in the BHM is not anymore negligible, but the atoms can still coherently tunnel from site to site. The interatomic interaction squeezes (i.e. reduces the fluctuations of) the number of atoms per site, $\Delta n \simeq G^{1/4}$. In this limit $\Delta n$ is smaller than the coherent state fluctuations ($\sim n^{1/2}$), yet is still greater than 1. Phases in different sites are correlated but their differences can have a significant dispersion which scales as

$$\Delta\phi \simeq \frac{1}{\Delta n} \simeq G^{-1/4} \ . \qquad (5)$$

- The phases become uncorrelated when $G \to G_c^+ \sim \mathcal{O}(1)$. At $G = G_c$ the system undergoes a quantum phase transition to an insulator (Mott) state characterized by a complete loss of phase coherence.

- $G < G_c$ —- The ground state of the BHM can be very well approximated by an incoherent superposition of localized states. At $G = 0$ the ground state becomes

$$|GS\rangle \ = \ |n, n, n, \ldots\rangle \ , \qquad (6)$$

The nature of the transition can be captured by a simple argument. On the one hand, in the state (6), an atom in each site attempts to hop with a rate $\gamma$. The total hopping energy of the $n$ atoms in each site is, therefore, $n\gamma$. On the other hand, to move one atom to a nearest neighbor site costs an interaction energy $\sim g\beta$. When $G < G_c = \mathcal{O}(1)$ the cost prevails over the gain provided by the hopping term, and the coherent tunneling is strongly suppressed. For any $G < G_c$ the dispersion of phase difference between nearest neighbor sites is maximal

$$\Delta\phi \ \sim \ 1 \ . \qquad (7)$$

Such simple energetic arguments cannot prove that the critical point $G = \mathcal{O}(1)$ describes a phase transition and not a crossover. We note, therefore, that the zero temperature model (1) in $d$ dimensions can be exactly mapped on the classical X-Y model at finite temperature in $d+1$ dimensions [12]. The $X-Y$ model has a continuous thermal phase transition. Thermal correlators in the X-Y model can be mapped back to quantum correlators in our quantum system. We will use this mapping in Section 1.6.

## 3.       The quantum phase model

For $G \ll n^2$ and $n \gg 1$ the boson Hubbard model (1) can be approximated by a quantum phase model (QPM) [15]:

$$\hat{H} = \sum_k \frac{g\beta}{2} \hat{n}_l^2 - \gamma \sum_{\langle k,l \rangle} \cos(\hat{\phi}_k - \hat{\phi}_l) \tag{8}$$

with the last sum running over the nearest-neighbor sites. The phase $\hat{\phi}_l$ and the number of atoms $\hat{n}_l$ in each site of the junction are (with some important caveats [16]) non-commuting conjugate observables $[\hat{n}_l, \hat{\phi}_l] = i$, and, in the $\phi$-representation,

$$\hat{n}_l \equiv n - i \frac{\partial}{\partial \phi_l} \ , \tag{9}$$

$$\hat{\phi}_l \equiv \phi_l \ . \tag{10}$$

Therefore, $\phi_l$ and $n_l$ play the role of coordinate and conjugate momentum, and satisfy the Heisenberg uncertainty relation. The insulator phase is characterized by large quantum phase fluctuations in the ground state, which destroy the long-range order among sites. The 1D Hamiltonian Eq.(8) exhibits a continuous Mott phase transition at the critical value $G_c = 0.617$ [12].

The QPM becomes equivalent to the BHM (1) in the limit of large number of atoms per site, $n \gg 1$, and for $G \ll n^2$ [17] (when $G \gg n^2$ the Schrœdinger equation contains extra $\cos 2(\phi_k - \phi_l)$ terms which have been neglected here). These conditions are well satisfied in current experiments where $\sim 10^5$ condensate atoms are trapped in $\sim 10^3$ wells created by a one dimensional optical lattice. Therefore, the QPM provides an unified model to study a nonequilibrium QPT in different systems, and, at the same time, allows the development of simple approximate methods and of a more transparent physical intuition of the process.

In what follows we study the microscopic dynamics of quantum phase transitions in a one-dimensional chain with periodic boundary conditions. The dynamics is governed by the QPM Eq.(8) which after rescaling time

$$t \ \rightarrow \ g\beta \, t \tag{11}$$

can be written as a dimensionless Schrœdinger equation

$$i\frac{\partial}{\partial t}\Psi = -\frac{1}{2}\sum_k \frac{\partial^2}{\partial \phi_k^2}\Psi - G \sum_{\langle k,l \rangle} \cos(\phi_k - \phi_l)\Psi \ . \tag{12}$$

The ground state below the transition ($G < G_c$) is close to the localized state (6) with all sites occupied by the same number of atoms $n$. In the phase representation this state is described by a uniform wavefunction, $\Psi(\phi_l) = \text{const}$, where all phase differences between nearest neighbor sites have the maximal dispersion of $\Delta\phi \sim 1$. Above the transition ($G > G_c$) the ground state is a number squeezed state which continuously tends to a Fock state when $G \to G_c^+$, and to a coherent state for $G \gg n^2$. Therefore, when $G \gg G_c$ and $G \ll n^2$ one can describe the low energy part of the spectrum of Eq.(12) in a harmonic approximation [18]

$$i\frac{\partial}{\partial t}\Psi = -\frac{1}{2}\sum_k \frac{\partial^2}{\partial\phi_k^2}\Psi + \frac{G}{2}\sum_{\langle k,l\rangle}(\phi_k - \phi_l)^2\Psi. \tag{13}$$

This equation is diagonalized by normal modes numbered by a lattice momentum

$$\mu \in \{-N_s + 1, \ldots, +N_s\}\ , \tag{14}$$

$\Psi = \prod_\mu \Psi_\mu(\Phi_\mu)$. Here $N_s$ is the number of lattice sites. There is one zero mode $\Phi_0 \sim \sum_l \phi_l$. All other modes $\Phi_\mu$ have nonzero frequencies $\sqrt{\gamma_\mu G}$,

$$i\frac{\partial}{\partial t}\Psi_\mu = -\frac{1}{2}\frac{\partial^2}{\partial\Phi_\mu^2}\Psi_\mu + \frac{\gamma_\mu G}{2}\Phi_\mu^2\Psi_\mu\ , \tag{15}$$

$$\gamma_\mu = 2\left[1 - \cos\left(\frac{\pi\mu}{N_s}\right)\right]\ . \tag{16}$$

We remark that all frequencies scale as

$$\sqrt{\gamma_\mu G}\ \sim\ G^{1/2}\ , \tag{17}$$

and phase dispersions in the ground states of the harmonic oscillators (15) scale as

$$\Delta\Phi_\mu\ =\ (\gamma_\mu G)^{-1/4}\ \sim\ G^{-1/4}\ . \tag{18}$$

We will use these scalings in the following.

## 4.     Linear Quench

To quench the system trough the transition point in a timescale $\tau_Q$, we linearly ramp the control parameter $G$ in Eq.(12) as

$$G(t)\ =\ \frac{t}{\tau_Q}\ . \tag{19}$$

The time $t$ runs from 0 to $t_{max} = \tau_Q G_{\max}$ with $G_{\max} \gg G_c$.

The initial state is the localized ground state at $G = 0$, see Eq.(6). The goal is to preserve memory of the phase disorder characteristic of the low $G$ Mott insulator phase into the large $G$ phase ordered superfluid phase. We begin with two technically trivial limits, $\tau_Q \to \infty$ and $\tau_Q \to 0$.

## 4.1 Adiabatic transition: $\tau_Q \to \infty$

In the limit $\tau_Q \to \infty$ the transition is adiabatic and the state of the system follows its instantaneous ground state from the initial localized state $|n, \ldots, n\rangle$ at $G = 0$ to a phase squeezed state with all the nearest neighbor phase differences close to zero, $\Delta \phi \approx 0$, at $G_{\max} \gg G_c$. The ground state at $G_{\max} \gg G_c$ is well described by the harmonic oscillators (15). Therefore, the final dispersion of phase differences in the adiabatic transition is

$$\Delta \phi_{\tau_Q \to \infty} \; \simeq \; G_{\max}^{-1/4} \; . \tag{20}$$

The phase dispersion in the final squeezed state can be probed with interference measurements.

## 4.2 Instantaneous transition: $\tau_Q \to 0$

In the opposite limit of an instantaneous transition, when $\tau_Q \to 0$, the system is not able to adjust its quantum state to the changing $G(t)$, and it remains in the initial localized state (6) till $G_{\max}$. The final dispersion of the phase differences in the uniform Fock state, $\Psi(\phi_l) = \text{const}$, is

$$\Delta \phi_{\tau_Q \to 0} \; \sim \; 1 \; . \tag{21}$$

The relative phases are random and we expect no interference fringes.

## 4.3 Diabatic transition

In the following we consider diabatic transitions in intermediate regimes. In Fig.(1) we illustratively show inverse gap $\tau = 1/dE$ of the system as a function of the control parameter $G(t)$. The inverse gap is a time scale on which the quantum system can react to changes of the external parameter $G(t)$. It plays the same role as the relaxation time in the thermal KZM. The inverse gap vanishes at the transition point $G = G_c$. For a generic quench the evolution of the system is approximately adiabatic when $G$ is far from $G_c$, and impulse when $G$ is close to $G_c$; compare with Fig.(2) where we consider a 3-site periodic lattice. This also happens in a thermal phase transition [2]. As $G(t)$ increases starting from 0, the system initially remains in its instantaneous ground state, which is an incoherent superposition of localized states, until the point $\hat{G}_- < G_c$ where the transition ceases to be adiabatic. Between $\hat{G}_-$ and a certain $\hat{G} > G_c$ the evolution is impulse: to a first approximation the state of the system does not change. At the point $G(t) = \hat{G}$ the system is still described by an incoherent superposition of localized states with a large phase dispersion $\Delta\phi \sim 1$. When $G(t) > \hat{G}$ the transition becomes adiabatic again.

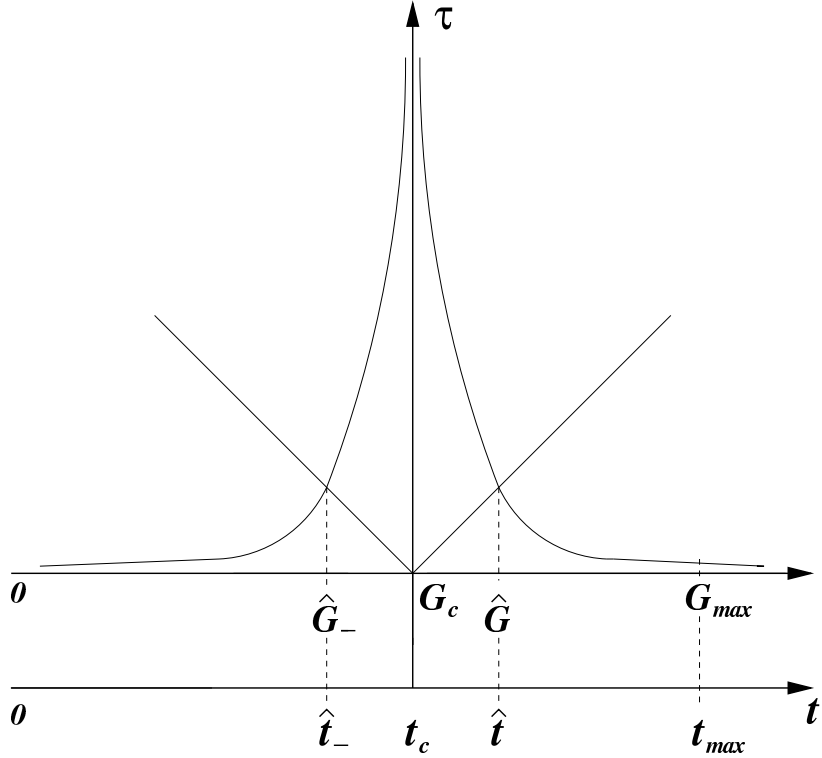

*Figure 1.* The inverse gap of the system as a function of the control parameter $G(t)$. The transition point is at $G_c = t_c/\tau_Q$. The transition is impulse when $\hat{G}_- < G < \hat{G}$. The time runs from 0 to $t_{max} = \tau_Q G_{max}$.

We consider two diabatic regimes

- the gaussian regime, which is reached when $G_c \ll \hat{G}$. In this regime we can use the gaussian approximation (15) when $G(t) \simeq \hat{G}$.

- the critical regime, reached when $\hat{G} \simeq G_c$. In this case the gaussian approximation (15) breaks down, and we need to use exact renormalization group critical exponents.

## 5. The gaussian regime

With the linear quench (19) the periods of the oscillators decay with time as:

$$\left(\frac{G(t)}{G_{max}}\right)^{-1/2} = \left(\frac{\tau_Q}{t}\right)^{1/2} . \qquad (22)$$

As shown in Appendix, the instantaneous time scale of the transition is

$$\frac{G}{\frac{dG}{dt}} = t \ . \tag{23}$$

This time scale becomes comparable to the periods of the short wavelength oscillators with $|\mu| \approx N_s$ at the time $\hat{t}$ when $G^{-1/2} \simeq G/\dot{G}$, see Fig.(2). The solution of this equation gives:

$$\hat{t} \simeq \tau_Q^{1/3} \ , \tag{24}$$

$$\hat{G} \equiv G(\hat{t}) \simeq \tau_Q^{-2/3} \ . \tag{25}$$

For $t > \hat{t}$ the short wavelength oscillators respond adiabatically to the quench. Oscillators with longer wavelengths become adiabatic later, but still at a time scaling as $\hat{t} \sim \tau_Q^{1/3}$. The system is in the gaussian regime, $G_c \ll \hat{G}$, when

$$\tau_Q \ll 1 \ . \tag{26}$$

The long wavelength oscillators become adiabatic at $\hat{G}_{|\mu|=1} \simeq (N_s/\tau_Q)^{-2/3}$. To remain in the gaussian regime, we want $\hat{G}_{|\mu|=1} \ll n^2$. This condition translates into

$$N_s \ll \tau_Q\, n^3 \ . \tag{27}$$

For, say, $n \approx 10^3$ atoms per site this condition is satisfied in a wide range of lattice sizes $N_s$ and quench times $\tau_Q$.

We now estimate the final dispersion $\Delta\phi$ of phase differences between the nearest neighbor lattice sites. At $\hat{G}$ the dispersion is $\Delta\phi \sim 1$ like in the initial Mott insulator state (6). For $G > \hat{G}$ the evolution of the harmonic oscillators is adiabatic. There is no mixing between the eigenstates of any oscillator $\mu$ (15). The dispersion of the phase $\Phi_\mu$ in any given eigenstate of the oscillator (15) scales as $G^{-1/4}$. In the adiabatic evolution after $\hat{G}$ the dispersion of $\Phi_\mu$ shrinks like $[\hat{G}/G(t)]^{1/4}$. $\Delta\phi$ behave in the same way. The final dispersion at $G_{\max}$ is

$$\Delta\phi \simeq \left(\frac{\hat{G}}{G_{\max}}\right)^{1/4} \simeq \tau_Q^{-1/6} G_{\max}^{-1/4} \ . \tag{28}$$

In a 1D periodic lattice this $\Delta\phi$ translates into a dispersion $\Delta L$ of the "angular momentum"

$$L = i \sum_l (a_l^\dagger a_{l+1} - a_{l+1}^\dagger a_l) \ , \tag{29}$$

which depends on the nearest neighbor $\Delta\phi$, through the formula

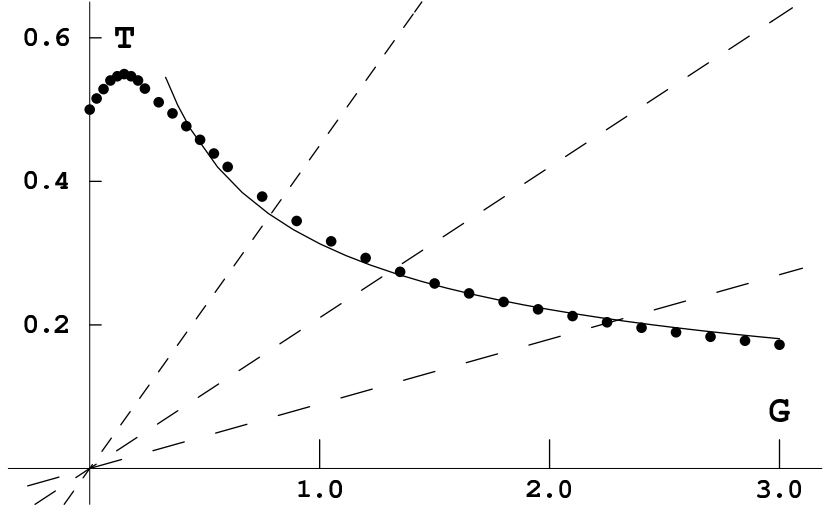

*Figure 2.* The inverse of the gap $T = 1/dE$ as a function of $G$ for a 3-site periodic Hubbard model with $n = 6$ atoms per site (dots). The solid line is the best $G^{-1/2}$ fit. The dashed lines show the transition time $G/\hat{G} = t = G\tau_Q/G_{\max}$ for $G_{\max} = 10$ and $\tau_Q = 4.5, 2.1, 0.9$ (from left to right). The crossings between $1/dE$ and the dashed lines define $\hat{G}$ for the different $\tau_Q$.

$$\Delta L \simeq \sqrt{N_s}\, n\, \Delta\phi \qquad (30)$$

valid for a small $\Delta\phi$. This $\Delta L$ may translate into a dispersion of the winding number $\sqrt{N_s}\Delta\phi$ when the atoms after the quench are somehow forced to condense.

Figures 3 and 4 show results of numerical simulations of the boson Hubbard model with a 3-site periodic lattice with a total of 300 atoms. In Fig.3 we show that $\Delta\phi$ remains the same as in the initial Fock state for $G < \hat{G}_{\tau_Q \ll 1}$. In Figure 4 we verify the scaling $\hat{t} \sim \tau_Q^{1/3}$.

## 5.1    Soft modes

In the argument of this Section we have neglected so far that different normal modes have different frequencies, see Eq.(16). We used only the general property that all frequencies, except for the zero mode, scale with $G$ as $G^{1/2}$. However, it is clear that long wavelength soft phonon modes will become adiabatic later than short wavelength modes. We can even define a $\hat{G}_\mu$ for each mode $\mu$ as a $G$ when the frequency of the mode becomes comparable to the transition rate,

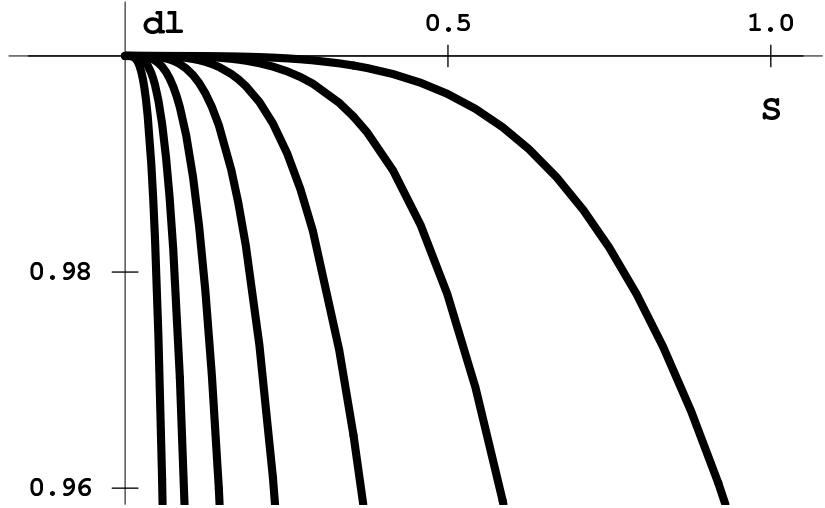

*Figure 3.* The normalized dispersion $dl = \Delta L / \Delta L_{\tau_Q \to 0}$ as a function of $S = G(t)/G_{\max}$. We show results of exact numerical simulations of the 3-site Hubbard model for 7 different $\tau_Q = 0.016, 0.008, 0.004, 0.002, 0.001, 0.0005, 0.00025$ (from left to right) with $G_{\max} = 1000$ and $n = 100$ atoms per site. $dl$ deviates from 1 (and hits the bottom of the figure) at $\hat{G} \sim \tau_Q^{-2/3}$: compare Fig.(4).

$$\sqrt{\gamma_\mu G} \simeq \frac{\dot{G}}{G} \, . \tag{31}$$

Solving this equation with respect to $G$ we obtain

$$\hat{G}_\mu \simeq \gamma_\mu^{-1/3} \, \tau_Q^{-2/3} \, . \tag{32}$$

After $\hat{G}_\mu$ the dispersion of the phase $\Phi_\mu$ shrinks like $G^{-1/4}$ and at a $G_{\max}$ it becomes

$$\Delta \Phi_\mu \simeq \left( \frac{\hat{G}_\mu}{G_{\max}} \right)^{1/4} \simeq \gamma_\mu^{-1/12} \tau_Q^{-1/6} G_{\max}^{-1/4} \, , \tag{33}$$

compare with Eq.(28). The dependence on $\gamma_\mu$ is much softened by the $-1/12$ exponent. $\gamma_\mu$ ranges from around $\pi^2/N_s^2$ for $\mu = 1$ to 2 for $\mu = N_s/2$, compare Eq.(16). For, say, $N_s = 100$ lattice sites in 1D the prefactor $\gamma_\mu^{-1/12}$ changes by less than a factor of 2. This justifies *a posteriori* the implicit approximation in our argument that $\gamma_\mu \simeq \mathcal{O}(1)$.

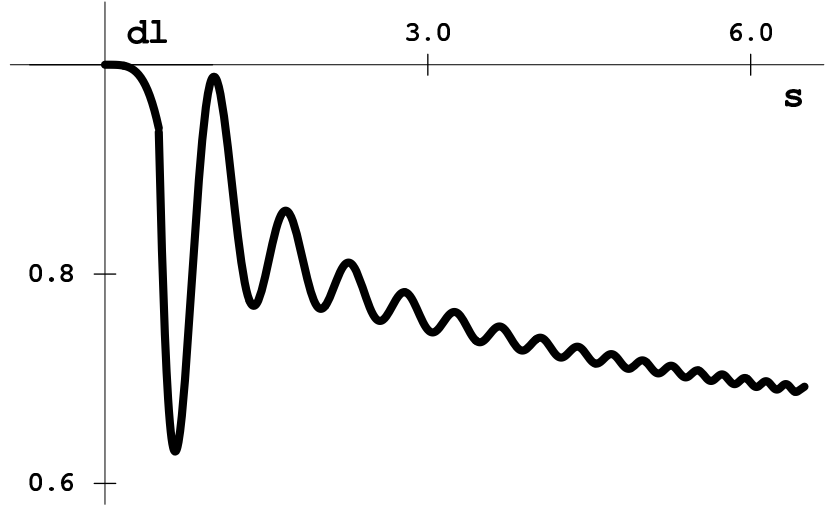

*Figure 4.* The normalized dispersion $dl = \Delta L/\Delta L_{\tau_Q \to 0}$ as a function of $s = G(t)/\hat{G}$ with $\hat{G} \sim \tau_Q^{-2/3}$. This figure displays all the seven plots from Fig.3. The plots in Fig.3 are in the top left corner of this figure. The fact that the seven plots sit on top of each other proves that the evolution of $\Delta L$ depends on time $t$ through the combination $t/\hat{t}$ with $\hat{t} \sim \tau_Q^{1/3}$.

## 5.2 Commensurate versus non-commensurate

So far we have assumed that the number of atoms $N$ is commensurate with the number of sites. In a rigorous mathematical sense, when $N$ is non-commensurate, there is no phase transition. The system is always in a superfluid phase. It is easy to see why. Suppose that we have $n$ atoms at each site and we add 1 atom to the system in order to make $N$ non-commensurate. The extra atom can hop between sites with a rate $\gamma$, since there is no interaction energy cost to suppress this hopping. The ground state for $G \ll 1$ is

$$|GS\rangle \sim \left(\sum_k a_k^\dagger\right) |n, n, n, \ldots\rangle . \tag{34}$$

The bulk of commensurate atoms are in an insulating state but the extra non-commensurate atom is superfluid even for this very small $G$.

In Fig.5 we show results of the same numerical simulation as in Fig.3 but with $N = 300 + 1$ atoms or $n = 100 + \frac{1}{3} \gg 1$ atoms per site. The plots in Fig.5 are essentially identical with the plots in Fig.3. This demonstrates that, at least in the gaussian regime, the transition is dominated by the commensurate bulk.

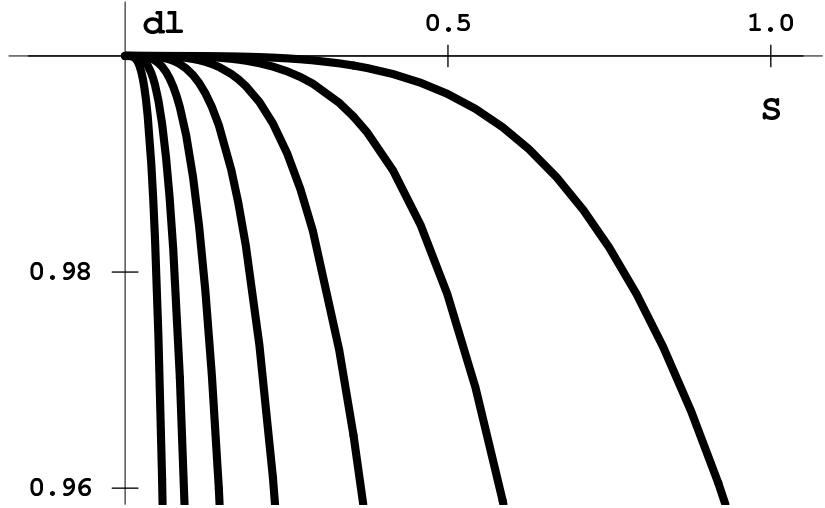

*Figure 5.* The normalized dispersion $dl = \Delta L/\Delta L_{\tau_Q \to 0}$ as a function of $S = G(t)/G_{max}$. We show results of exact numerical simulations of the 3-site Hubbard model for 7 different $\tau_Q = 0.016, 0.008, 0.004, 0.002, 0.001, 0.0005, 0.00025$ (from left to right) with $G_{max} = 1000$ and $n = 100 + \frac{1}{3}$ atoms per site. These plots are indistinguishable from plots in Fig.3 obtained for an integer $n = 100$.

## 6. The critical regime

So far we have considered the case $\tau_Q \ll 1$ with $\hat{G}$ well inside the harmonic regime. In slower transitions, when $\hat{G} \simeq G_c$, the harmonic approximation cannot be applied. The slow transitions ($\tau_Q \ll 1$) probe the critical behavior of the system close to $G_c$.

- In the periodic 3-site array there is no phase transition but, rather, a crossover: the gap $dE$ is minimal at the crossover point $G_c \approx 0.5$ but it does not vanish there, compare Figs.(1,2). For $\tau_Q \gg 1$ the transition is adiabatic and the system follows its instantaneous ground state. The phase difference dispersion at $G_{max} \gg G_c$ is the dispersion in the ground state of the harmonic oscillators (20). A thermal crossover transition was studied in [21].

- In a large 1 D array with a commensurate density of atoms there is a phase transition and not a crossover [12, 19]. The gap $dE$ vanishes at $G = G_c$. The quantum zero temperature transition translates into Berezinskii-Kosterlitz-Thouless transition which does not have a well defined correlation length in the ordered phase - correlations decay alge-braically with a distance on a lattice. The scaling argument alone is not enough to make predictions of $\Delta\phi$ in this case.

■ In a large 2 D array with a commensurate density of atoms there is a second order phase transition. Close to the critical point, $G \approx G_c$, the energy gap scales as

$$dE \sim (G - G_c)^\nu \,, \tag{35}$$

where $\nu = 2/3$ is an exact renormalization group critical exponent of the X-Y model in $2 + 1$ D. [1] With a linearized

$$G(t) - G_c \;=\; \frac{t}{\tau_Q} \tag{36}$$

the gap $\Delta E \sim (G - G_c)^{2/3} = (t/\tau_Q)^{2/3}$ becomes equal to the quench rate $\dot{G}/(G - G_c) = 1/t$ at

$$\hat{G} \;\simeq\; G_c \;+\; \frac{\mathcal{O}(1)}{\tau_Q^{3/5}} \,. \tag{37}$$

Before $\hat{G}$ the system is in an incoherent superposition of localized states. After $\hat{G}$ the evolution is adiabatic, and with increasing $G$ the phase dispersion shrinks together with the phase widths of the system eigenstates. The dispersion measured at $G_{\max} \gg G_c$, where we can use the harmonic approximation (13,18), is

$$\Delta\phi_{\tau_Q \gg 1}^{\mathrm{2D}} \;\approx\; \frac{\hat{G}^{1/4}}{G_{\max}^{1/4}} \;\simeq\; \frac{\left(G_c + \frac{\mathcal{O}(1)}{\tau_Q^{3/5}}\right)^{1/4}}{G_{\max}^{1/4}} \,. \tag{38}$$

This formula is consistent with Eq.(20) because $G_c = \mathcal{O}(1)$. Here we again use the fact that the phase dispersion of harmonic oscillator eigenstates shrinks like $G^{-1/4}$. The critical behavior is more transparent in a phase dispersion offset with a phase dispersion for large $\tau_Q$,

$$\Delta\phi_{\tau_Q \gg 1}^{\mathrm{2D}} - \Delta\phi_{\tau_Q \to \infty} \;\simeq\; \frac{1}{G_{\max}^{1/4}\, \tau_Q^{3/5}} \,. \tag{39}$$

The critical exponent $\nu = 2/3$ determines the exponent of $3/5$.

■ In a large 3 D array there is a second order phase transition with a gaussian critical exponent of $\nu = 1/2$ characteristic for the X-Y model in $3 + 1$ D. In 3 D the results of the gaussian regime are valid also in the critical regime. The Mott transition has been recently observed in 1 D [22], and in 3 D [23] .

## 7.    From superfluid to insulating phase

So far we have focused on transitions from the insulating to the superfluid phase where the non-equilibrium dynamics of the transition leaves imprint on the dispersion of phase difference between sites $\Delta\phi$. The phase dispersion can be measured by interference techniques [20]. In this Section we will reverse the direction of the transition to go from the superfluid to the insulating phase. We will use the dispersion in the number of atoms per site $\Delta n$ as a stable record of the nonequilibrium dynamics of the transition. In the insulating phase the number of atoms in each site is conserved.

Again we linearly ramp $G$ from a $G_{max}$ ( $1 \ll G_{max} \ll n^2$) at $t = -\tau_Q G_{max}$ down to $G = 0$ at $t = 0$,

$$G(t) \, = \, \frac{t}{\tau_Q} \, . \qquad (40)$$

We can distinguish several regimes for different values of $\tau_Q$.

- **Instantaneous transition with $\tau_Q \to 0$.** The system remains in the initial ground state at $G_{max}$ which is a squeezed quasi-coherent state with a dispersion

$$\Delta n_{\tau_Q \to 0} \, \simeq \, G_{max}^{1/4} \, . \qquad (41)$$

- **Adiabatic limit of $\tau_Q \to \infty$.** The state of the system follows the instantaneous ground state. The system ends in the Mott insulator state (6) with a dispersion

$$\Delta n_{\tau_Q \to \infty} \, = \, 0 \, . \qquad (42)$$

- **Diabatic transition with $\tau_Q \ll 1$ - the gaussian regime.** The characteristic frequency in the harmonic regime scales like $G^{1/2}$, compare Eq.(17). This frequency becomes equal to the transition rate $\dot{G}/G$ at a $\hat{t}$ when $\hat{G} \simeq \tau_Q^{-2/3}$. Before $\hat{t}$ the system follows its ground state which is a squeezed state with $\Delta n \simeq \hat{G}^{1/4}$. At $\hat{t}$ when $\Delta n \simeq \hat{G}^{1/4}$ the evolution becomes impulse, the state does not change with the decreasing $G(t)$ any more. The system arrives at $G = 0$ with

$$\Delta n_{\tau_Q \ll 1} \, \simeq \, \hat{G}^{1/4} \, \simeq \, \tau_Q^{-1/6} \, . \qquad (43)$$

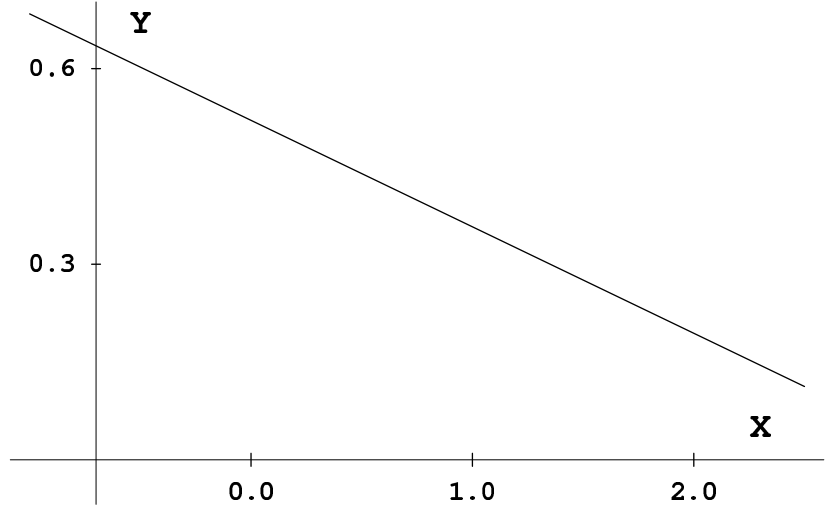

*Figure 6.*    Logarithm of the dispersion of the number of atoms per site $Y = \log_{10} \Delta n$ as a function of $X = \log_{10} \tau_Q$. The dots are averages over many numerical simulations of the 3-site periodic Hubbard model (1) for $n = 100$ atoms per site. The solid line is the best linear fit with a slope of $-0.164 \pm 0.002$. This slope compares well with the prediction of $-1/6 = -0.167$ in Eq.(43).

This scaling applies when $\hat{G} \ll G_{\max}$ or $\tau_Q \gg G_{\max}^{-3/2}$. In the opposite case the transition is instantaneous.

In Fig.6 we show $\Delta n$ as a function of $\tau_Q$ in a 3-site periodic Hubbard model with $n = 100$ atoms per site. The best fit to the numerical results gives an exponent of $-0.164 \pm 0.002$ which is consistent with the prediction of $-1/6 = 0.167$ in Eq.(43).

- **Diabatic transition with $\tau_Q \gg 1$ - the critical regime.** For a 2D or 3D array the gap $dE$ depends on a distance from the critical point $|G_c|$ as $dE \sim |G - G_c|^\nu$ with $\nu = \frac{2}{3}$ in 2D and $\nu = \frac{1}{2}$ in 3D. The dispersion $\Delta n$ in the ground state at $G_c$ is $\Delta n_c \approx 0.5$ and close to $G_c$ the dispersion can be linearized as $\Delta n - \Delta n_c \simeq G - G_c = -t/\tau_Q$. The evolution is impulse between two times, $-\hat{t}$ and $+\hat{t}$, when the transition rate $\dot{G}/(G - G_c) = 1/|t|$ equals the gap $dE \sim |G - G_c|^\nu$. This happens when $|\hat{G} - G_c| \simeq \tau_Q^{-1/(1+\nu)}$. The system adiabatically follows its instantaneous ground state until $-\hat{t}$ when the dispersion in the ground state is $\Delta n_- - \Delta n_c \simeq +|\hat{G} - G_c|$. Between $-\hat{t}$ and $+\hat{t}$ the evolution is impulse, in the first approximation the state of the system does not change. The dispersion in the instantaneous ground state at $+\hat{t}$ is $\Delta n_+ - \Delta n_c \simeq -|\hat{G} - G_c|$ but the state of the system still has the dispersion $\Delta n_- > \Delta n_+$. This extra dispersion above that in the instantaneous ground state is

$$\Delta n_- \, - \, \Delta n_+ \, \simeq \, \tau_Q^{-\frac{1}{1+\nu}} \; . \tag{44}$$

This is dispersion at $+\hat{t}$ but we want to know the final dispersion when $G = 0$.

In the insulating phase the $\Delta n$ of the ground state strongly depends on $G$, it ranges from $\Delta n = 0$ at $G = 0$ to $\Delta n_c \approx 0.5$ at $G = G_c \approx 1$. In contrast, $\Delta n$ of excited states remains nonzero even at $G = 0^+$ because an arbitrarily small but nonzero tunneling rate $G$ removes degeneracy of states with different occupation numbers. All excited states at $G = 0^+$ have nonzero $\Delta n$. Their $\Delta n$ depends on $G$ but, in first approximation, this dependence can be neglected as compared to that in the ground state. In short, after $+\hat{t}$ the $\Delta n$ of the ground state shrinks down to 0 while the $\Delta n$'s of the excited states does not change. The state of the system at $t = +\hat{t}$ is a superposition of the ground state and the excited states. The adiabatic evolution after $+\hat{t}$ ends in a state with a $\Delta n$ that is a linear function of the difference in Eq.(44),

$$\Delta n_{\tau_Q \gg 1} \, \simeq \, \tau_Q^{-\frac{1}{1+\nu}} \; . \tag{45}$$

This final $\Delta n$ depends both on the transition time $\tau_Q$ and on the critical exponent $\nu$.

## 8. Conclusion

After crossing the transition, the system does not have a definite angular momentum or definite phase differences between lattice sites, but it is in a quantum superposition of states with different $\Delta\phi$. Either a measurement or decoherence [24] are needed to convert this coherent superposition into a mixture of states, each with definite current. This is an important difference with respect to the case of a thermal KZM, where $\Delta\phi$ would describe the dispersion in an ensemble of different possible classical outcomes. Moreover, the quantum dynamics is reversible, while in the thermal case the characteristic lenghtscale $\hat{\xi}$ is frozen after the symmetry breaking transition is completed. $\hat{\xi}$ is a record of the transition rate $\tau_Q$: manipulations with $\epsilon$ do not change the winding number as long as the system remains in the symmetry broken phase with $\epsilon > 0$. In contrast, even after the diabatic quantum transition is completed one can change $\Delta\phi_{\tau_Q}$. Adiabatic variations of $G$ away from $G_{\max}$ are accompanied by changes in $\Delta\phi$,

$$\Delta\phi = \Delta\phi_{\tau_Q}(G_{\max}/G)^{1/4} \; . \tag{46}$$

In conclusion, we have predicted the phase dispersion after a diabatic insulator-superfluid quantum phase transition in an array of weakly coupled Bose-Einstein condensates or Josephson junctions. This theory is a quantum counterpart of the Kibble-Zurek mechanism for topological defect formation in classical thermal phase transitions. Our predictions can be tested experimentally with a Bose-Einstein condensate trapped in a periodic potential [20, 23]. In this case our Eq.(28) predicts a dispersion of the phases $\Delta\phi \simeq G_{\max}^{-1/4} \tau_Q^{-1/6}$ for a transition to the superfluid phase and a dispersion of the number of atoms per lattice site of $\Delta n \sim \tau_Q^{-1/6}$ for a transition to the insulating phase. The dispersions can be directly measured by, respectively, interference and phase contrast imaging techniques.

## Acknowledgments

This work was partially supported by DOE. J.D. was supported in part by the KBN grant 2 P03B 092 23.

## Appendix

Here we justify the choice of $\frac{dG}{dt}/G$, and not, say, $\frac{dG}{dt} = \tau_Q^{-1}$, as the transition rate. The wavefunction in Eq.(15) can be expanded in eigenstates of the harmonic oscillator,

$$\Psi_\mu[\Phi_\mu] = \sum_{m=0}^{\infty} A_m(t)\, U_m \left[ \frac{\Phi_\mu}{(\gamma_\mu G)^{1/4}} \right], \qquad (A.1)$$

where $U_m[x]$ are eigenstates of a dimensionless harmonic oscillator

$$-\frac{1}{2}\frac{d^2 U_m[x]}{dx^2} + \frac{1}{2}x^2 U_m[x] = \left(m + \frac{1}{2}\right) U_m[x]. \qquad (A.2)$$

Substitution of the expansion (A.1) to the Schrödinger equation (15) gives

$$i \sum_{m=0}^{\infty} \frac{dA_m}{dt} U_m =$$
$$\sum_{m=0}^{\infty} A_m \left[ \left(\sqrt{\gamma_\mu G}\right) \left(m + \frac{1}{2}\right) U_m - \frac{i}{4} x \left(\frac{\frac{dG}{dt}}{G}\right) \frac{dU_m}{dx} \right].$$

Here $x = \Phi_\mu/(\gamma_\mu G)^{1/4}$. We can eliminate the first term on the right hand side by a redefinition of the amplitude

$$A_m(t) = B_m(t) \exp\left(-i \int_0^t dt' \sqrt{\gamma_\mu G(t')}\right). \qquad (A.3)$$

This is just a change of phase factors. After projection on $U_n[x]$ we get

$$\frac{dB_n}{dt} = -\frac{1}{4} \left(\frac{\frac{dG}{dt}}{G}\right) \times$$

$$\sum_{m=0}^{\infty} B_m e^{i(n-m)\int_0^t dt' \sqrt{\gamma_\mu G(t')}} \int dx \, x U_n \frac{dU_m}{dx} \; . \qquad (A.4)$$

This equation describes flow of amplitudes between different modes. The $\left(\frac{\frac{dG}{dt}}{G}\right)$ sets the rate at which the $B_n$ are forced to change. However, the right hand side oscillates with frequencies which are multiplicities of the instantaneous frequency of the mode $\sqrt{\gamma_\mu G(t)}$. The effect of the right hand side on $B_n$ averages to zero when

$$\left(\frac{\frac{dG}{dt}}{G}\right) \ll \sqrt{\gamma_\mu G(t)} \; . \qquad (A.5)$$

With this condition different amplitudes $B_n$ effectively decouple and the evolution is adiabatic.

# References

[1] T.W.B. Kibble, J. Phys. **A9**, 1387 (1976); T.W.B. Kible and A. Vilenkin, Phys. Rev. **D52**, 679 (1995).

[2] W.H. Zurek, Nature (London) **317**, 505 (1985).

[3] W.H. Zurek, Phys. Rep. **276**, 177 (1996).

[4] M. Hindmarsh and A. Rajantie, Phys.Rev.Lett.**85**, 4660 (2000); G.J. Stephens, L.M.A. Bettencourt, and W.H. Zurek, cond-mat/0108127.

[5] P.C. Hendry *et.al.*, Nature **368**, 315–317 (1996); M.E. Dodd *et.al.*, Phys.Rev.Lett. **81**, 3703–3706 (1998).

[6] V.M.H. Ruutu et al, Nature **382**, 334 (1996); Bauerle *et.al.*, Nature **382**, 332 (1996).

[7] P. Laguna and W.H. Zurek Phys.Rev.Lett. **78**, 2519-2522 (1997); A. Yates and W.H. Zurek, Phys.Rev.Lett. **80**, 5477-5480 (1998); N.D. Antunes, L.M.A. Bettencourt, and W.H. Zurek, Phys.Rev.Lett. **82**, 2824-2827 (1999).

[8] G.D. Lythe, Phys.Rev. **E 53**, R4271-R4274 (1996); E. Moro and G. Lythe, Phys.Rev. **E 59**, R1303-R1306 (1999).

[9] R. Carmi *et al.*, Phys. Rev. Lett.**84**, 4966 (2000); E. Kavoussanaki *et al.*, Phys. Rev. Lett.**85**, 3452 (2000).

[10] R. Monaco, J. Mygind, and R.J. Rivers, Phys.Rev.Lett. **89**, 080603 (2002).

[11] J.R. Anglin and W.H. Zurek, Phys.Rev.Lett. **83**, 1707 (1999).

[12] S. Sachdev, Quantum Phase Transitions, Cambridge University Press, (1999).

[13] J. Dziarmaga, A. Smerzi, W.H. Zurek and A.R. Bishop, Phys.Rev.Lett.**88**, 167001 (2002).

[14] D. Jaksch et al., Phys. Rev. Lett. **81**, 1322 (1998)

[15] E. Simanek, *Inhomogeneous Superconductors*, Oxford University Press, 1994.

[16] J. Anglin, P. Drummond and A. Smerzi, Phys.Rev **A64**, 063605 (2001)

[17] With this assumption we could neglect in Eq.(2) the $O(G/n^2)$ correction terms derived in [16].

[18] J. Javanainen, Phys. Rev. **A60**, 4902 (1999).

[19] M.P.A. Fisher, P.B. Weichman, G. Grinstein, and D.S. Fisher, Phys.Rev.**B** 40, 546 (1989).

[20] B.P. Anderson and M.A. Kasevich, Science **282**, 1686 (1998); C. Orzel et al., Science **291**, 2386 (2001); F.S. Cataliotti *et.al.*, Science **293**, 843 (2001).

[21] J. Dziarmaga, Phys.Rev.Lett.**81**, 5485 (1998); W.H. Zurek, L.M.A. Bettencourt, J. Dziarmaga, and N.D. Antunes, Acta.Phys.Polon.**B** 32, 2279 (2000).

[22] M.A. Kasevich et al., Yale University, private communication.

[23]  M. Greiner *et al.*, Nature **415**, 39 (2002).

[24]  W.H. Zurek, Phys. Today **44**, 36 (1991); quant-ph/010527.